\documentclass[11pt,preprint,tightenlines,
showpacs,preprintnumbers,amsmath,amssymb,
a4paper,nofootinbib]{revtex4-1}

\usepackage{nicefrac}

\usepackage{fouriernc}

\usepackage{natbib}
\usepackage{amssymb,amsbsy,amsmath,amsfonts}
\usepackage{graphicx}
\usepackage{epsf,epsfig,float,latexsym,amsthm,fancyhdr,rotating}
\usepackage{graphics,psfrag,longtable}
\usepackage{slashed}
\usepackage[colorlinks,citecolor=blue,linktoc=all,linkcolor=cyan,urlcolor=blue]{hyperref}
\usepackage{upgreek}

 \usepackage{setspace}

\def\beq{\begin{equation}}
\def\eeq{\end{equation}}
\def\bea{\begin{eqnarray}}
\def\eea{\end{eqnarray}}
\def\beqa{\begin{equation}\begin{array}{l}}
\def\eeqa{\end{array}\end{equation}}
\def\eqlab#1{\label{eq:#1}}

\def\Eqref#1{Eq.~(\ref{eq:#1})}

\def\Figref#1{Fig.~\ref{fig:#1}}

\def\half{\mbox{$\frac{1}{2}$}}

\def\barr{\left(\begin{array}{c}}
\def\earr{\end{array}\right)}
\def\bmat{\left(\begin{array}{cc}}
\def\emat{\end{array}\right)}
\def\al{\alpha}
\def\be{\beta}

\def\nn{\nonumber}
\def\dd{\mathrm{d}}

\def\3d{3-D}

\def\ol#1{\overline{#1}}

\def\bq{\mathbf{q}}

\begin{document}
\title { The subtraction contribution to the muonic-hydrogen Lamb shift:
a point for lattice QCD calculations of the polarizability effect}

\author{Franziska Hagelstein}
\affiliation{Paul Scherrer Institut, CH-5232 Villigen PSI, Switzerland}

\author{Vladimir Pascalutsa}
\affiliation{Institut f\"ur Kernphysik,
 Johannes Gutenberg-Universit\"at  Mainz,  D-55128 Mainz, Germany}

\begin{abstract}
 The proton-polarizability contribution to the
 muonic-hydrogen Lamb shift is a major source of 
 theoretical uncertainty in the extraction of the proton charge radius. An empirical evaluation of this effect, based on the proton structure functions, requires a systematically improvable 
 calculation of the ``subtraction function'', possibly using lattice QCD. 
 We consider a different subtraction point, with the aim of accessing the subtraction function directly in lattice calculations. A useful feature of this subtraction point is that the corresponding contribution of the structure functions to the Lamb shift is suppressed. The whole effect is dominated by the
 subtraction contribution, calculable on the lattice.
\end{abstract}
\date{\today}
\maketitle

\tableofcontents

\section{Introduction --- The point of subtraction}

The nucleon structure functions are used as input
in calculations of the nuclear structure effects in precision atomic spectroscopy (see \cite{Pohl:2013yb,Carlson:2015jba,Hagelstein:2015egb,Pasquini:2018wbl} for reviews).
The muonic-hydrogen ($\mu$H) experiments demand a higher quality of this input
in both the Lamb shift \cite{Pohl:2010zza,Antognini:1900ns} and hyperfine structure \cite{Pohl:2016xsr,Bakalov:2015xya,Kanda:2018oay}. 
In the Lamb-shift calculations, however, the problem is severed by the fact that, in addition
to the unpolarized  structure functions $F_{1}(x, Q^2)$ and $F_{2}(x, Q^2)$, one encounters a ``subtraction function'' which is largely unknown. Thus, the subtraction-function contribution to the Lamb shift 
precludes insofar any systematic improvement of the theoretical uncertainty of 
the ``data-driven'' evaluations, based on dispersion relations alone. 

Ultimately, one should aim at calculating the subtraction-function contribution
in lattice QCD (LQCD). Some efforts in this direction have already been made \cite{Ji:2001wha,Chambers:2017dov, Can:2020,Hannaford-Gunn:2020pvu} 
 In the present work we propose  an unconventional choice of the subtraction point, which, first of all, is directly accessible in lattice calculations, and secondly, diminishes the structure-function contribution.  
\begin{figure}[b]
\centering
       \includegraphics[width=5.5cm]{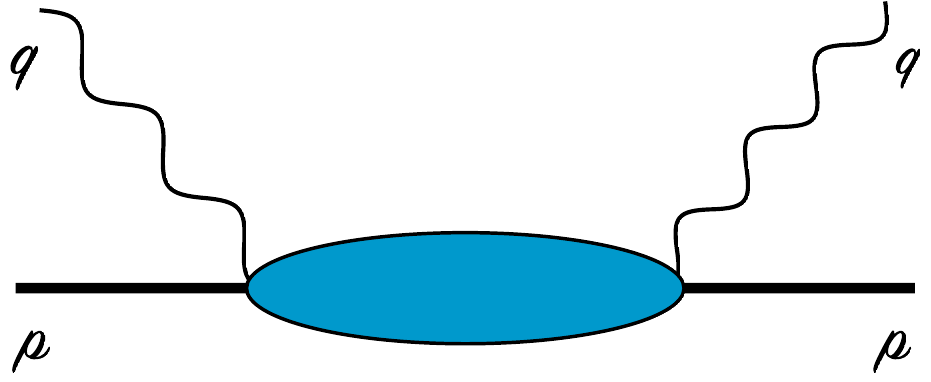}
\caption{Forward doubly-virtual Compton scattering
process.  \label{fig:CSgeneric}}
\end{figure}

We start by considering 
the forward doubly-virtual Compton scattering (VVCS) amplitude (cf.\ \Figref{CSgeneric}): 
\beq
T^{\mu\nu} (p,q) = i\int d^4 x\, e^{-i q\cdot x}\, \big\langle N(p)\big| T \big( j^\mu(x)\, j^\nu (0) \big) \big| N(p) \big\rangle , 
\eeq
with $p$ and $q$ the target and photon four-momenta, respectively.
For an unpolarized target (of any spin), 
the forward VVCS amplitude decomposes into two Lorentz structures:
\beq 
T^{\mu\nu} (p,q) = \left( -g^{\mu\nu}+\frac{q^{\mu}q^{\nu}}{q^2}\right)
T_1(\nu, Q^2) +\frac{1}{M^2} \left(p^{\mu}-\frac{p\cdot
q}{q^2}\,q^{\mu}\right) \left(p^{\nu}-\frac{p\cdot
q}{q^2}\, q^{\nu} \right) T_2(\nu, Q^2),\eqlab{fVVCS}
\eeq
where $\sqrt{p^2} = M$ is the target mass.
The two scalar amplitudes $T_{1,2}$ are functions
of the photon energy $\nu = p\cdot q /M$ and the 
virtuality $Q^2=-q^2$. 

\begin{figure}[t]
\centering
       \includegraphics[width=5.5cm]{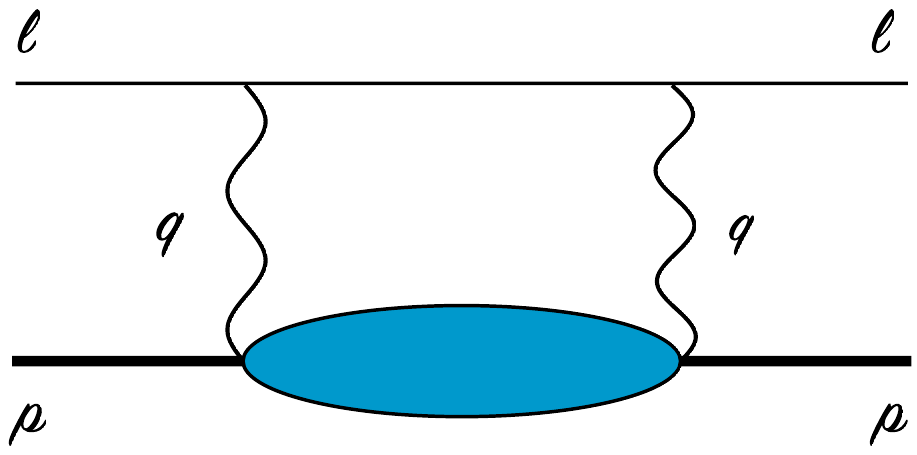}
\caption{Forward two-photon exchange in a hydrogen-like atom.  \label{fig:TPE}}
\end{figure}

The two-photon exchange (TPE) correction to the $n$-th $S$-level of a hydrogen-like atom, shown in Fig.\ \ref{fig:TPE}, is subleading to the well-known proton charge radius effect. To order $\al^5$, the TPE contribution is expressed entirely in terms of the forward VVCS amplitudes:
\beq
\Delta E_{nS}= 8\pi \al m \,\phi_n^2\,
\frac{1}{i}\int_{-\infty}^\infty \!\frac{\dd\nu}{2\pi} \int \!\!\frac{\dd \bq}{(2\pi)^3}   \frac{\left(Q^2-2\nu^2\right)T_1(\nu,Q^2)-(Q^2+\nu^2)\,T_2(\nu,Q^2)}{Q^4(Q^4-4m^2\nu^2)},\eqlab{VVCS_LS}
\eeq
where $\al$ is the fine-structure constant, $\phi_n^2=1/(\pi n^3 a^3)$ is the Coulomb wave-function at the origin, $a=(Z \alpha m_r)^{-1}$ is the Bohr radius, $m_r=m M /(m+M)$ is the reduced mass, $M$ and $m$ are the nucleus and lepton masses, and $Z$ is the charge of the nucleus (in the following we assume $Z=1$).

To date, all the data-driven evaluations of the TPE effect in $\mu$H,
employ the following dispersion relations:
\begin{subequations}
\eqlab{T1T2DRelinel}
\bea
T_1 ( \nu, Q^2) &=& T_1(0,Q^2) +\frac{32\pi \al M\nu^2}{Q^4}\int_{0}^1 
\,\dd x \, 
\frac{x\, F_1 (x, Q^2)}{1 - x^2 (\nu/\nu_{\mathrm{el}})^2 - i 0^+}\eqlab{T1Subtr}, \\
T_2 ( \nu, Q^2) &=& \frac{16\pi\al M}{Q^2} \int_{0}^1 
\!\dd x\, 
\frac{F_2 (x, Q^2)}{1 - x^2 (\nu/\nu_{\mathrm{el}})^2  - i 0^+}, \eqlab{T2DR}
\eea
\end{subequations}
with $\nu_\mathrm{el}=\nicefrac{Q^2}{2M}$ and the Bjorken variable $x$. 
Note that, while $T_2$ is fully determined by the empirical proton structure function $F_2(x,Q^2)$, 
the amplitude $T_1$ is determined by $F_1(x,Q^2)$ only up to the subtraction function at $\nu=0$. 
This is because the high-energy behaviour of the proton structure function $F_1$ requires at least one subtraction
for a valid dispersion relation.

The point of subtraction, however,  is rather arbitrary. Recently, 
Gasser, Leutwyler and Rusetsky have considered the subtraction at $\nu = \half i Q$ and argued that it is advantageous in the context of the Cottingham formula for the proton-neutron mass splitting \cite{Gasser:2020mzy}. Here, we find that the subtraction at $\nu = iQ$ leads to
interesting ramifications for the $\mu$H Lamb-shift evaluation.
In this case, for example,
the so-called ``inelastic contribution'' (i.e., 
the contributions of inelastic structure functions $F_1$ and $F_2$)
becomes negligible. In what follows, we go through the main steps of the formulae, present numerical results based on the Bosted-Christy parametrization of proton structure functions \cite{Christy:2011}, and discuss the prospects for future lattice calculations of these effects.

\section{Standard formulae}
Recall that the VVCS amplitudes can be split unambiguously into Born and polarizability pieces:
\beq
T_i  = T_i^{\mathrm{Born}} + \ol T_i,  
\eeq 
with the Born term given
entirely by the elastic form factors $G_E(Q^2)$ and
$G_M(Q^2)$, whereas the remainder $\ol T_i $ is,
in the low-energy limit,
characterised by polarizabilities, e.g.:
\begin{subequations}
\bea 
\ol T_1 (\nu, Q^2) &=& 4\pi \left[ \beta_{M1} Q^2 + \big(\al_{E1}+\be_{M1}\big) \nu^2  \right] + O(Q^4,\nu^4, Q^2\nu^2) \\ 
\ol T_2 (\nu, Q^2) &=& 4\pi  \big(\al_{E1}+\be_{M1}\big) Q^2   + O(Q^4,\nu^4, Q^2\nu^2),
\eea 
\end{subequations}
where $\al_{E1}$ ($\be_{M1}$) is the electric (magnetic) dipole polarizability. In general, the polarizability term by
itself satisfies the aforementioned dispersion relations (omitting the $i0^+$ prescription from now on) :
\begin{subequations}
\bea
\ol T_1 ( \nu, Q^2) &=& \ol T_1(0,Q^2) +\frac{32\pi \al M\nu^2}{Q^4}\int_{0}^{x_0} 
\,\dd x \, 
\frac{x\, F_1 (x, Q^2)}{1 - x^2 (\nu/\nu_{\mathrm{el}})^2 }\eqlab{olT1Subtr}, \\
\ol T_2 ( \nu, Q^2) &=& \frac{16\pi\al M}{Q^2} \int_{0}^{x_0}
\!\dd x\, 
\frac{F_2 (x, Q^2)}{1 - x^2 (\nu/\nu_{\mathrm{el}})^2 }. \eqlab{olT2DR}
\eea
\end{subequations}
The only distinction with the above \Eqref{T1T2DRelinel} is the upper limit of integration over the Bjorken-$x$, which is now set
by an inelastic threshold $x_0$, thus excluding explicitly the elastic contribution. 

The full TPE calculation, in the data-driven approach based
on dispersion relations, is then split into three contribution:
elastic, inelastic and subtraction. The elastic one follows
directly from the Born contribution to VVCS amplitudes and it yields essentially the effect of the Friar radius (a.k.a., the 3\textsuperscript{rd} Zemach moment). The other two contributions follow from the polarizability
piece, such that the inelastic contribution
is given by the inelastic structure functions, whereas 
subtraction is the rest, i.e., the $nS$ level is affected as:
\begin{subequations}
\eqlab{standard}
\bea
\Delta E^{(\mathrm{inel.})}_{nS}&=&-32\al^2Mm\,\phi_n^2\,\int_0^\infty \frac{\dd Q}{Q^5}\,\int_0^{x_0}\frac{\dd x}{(1+v_l)(1+v_x)}\Bigg\{\left[1+\frac{v_lv_x}{v_l+v_x}\right]F_2(x,Q^2)\nn\\
&&+\frac{2x}{(1+v_l)(1+v_x)}\left[2+\frac{3+v_lv_x}{v_l+v_x}\right]F_1(x,Q^2)\Bigg\},\eqlab{inelasticpol}\\
\Delta E^{\mathrm{(subtr.)}}_{nS}&=&\frac{2\al m}{\pi}\,\phi_n^2\,\int_0^\infty \frac{\dd Q}{Q^3}\frac{2+v_l}{(1+v_l)^2}\,  \ol T_1(0,Q^2),\eqlab{subpol}
\eea
\end{subequations}
with the following definitions:
\begin{subequations}
\bea
 v_x&=&\sqrt{1+x^2\tau^{-1}} ,  \qquad \tau=\frac{Q^2}{4M^2},\\ 
 v_l&=&\sqrt{1+\tau_l^{-1} },  \qquad
\tau_l=\frac{Q^2}{4m^2}.
\eea
\end{subequations}
These are the standard formulae, where the subtraction is made at $\nu=0$. 

\section{A different subtraction point}
Now, for the subtraction at $\nu = iQ $, the subtracted
dispersion relation reads as:
\beq
\ol T_1 ( \nu, Q^2) = \ol T_1(iQ,Q^2)+\frac{32 \pi \al M}{Q^4}\left(\nu^2+Q^2\right)\int_0^{x_0} \dd x\,x \,\frac{F_1(x,Q^2)}{\big[1-x^2 (\nu/\nu_\mathrm{el})^2\big]
\big(1+x^2 \tau^{-1} \big)}.
\eeq
Hence, the inelastic and subtraction contributions change into:
\begin{subequations}
\bea
\Delta  E^{\prime \, (\mathrm{inel.})}_{nS} &=& -32\al^2Mm \phi_n^2 \int_0^\infty \frac{\dd Q}{Q^5}\int_0^{x_0}\ \dd x\,\Bigg\{\frac{1}{(1+v_l)(1+v_x)}\left[1+\frac{v_lv_x}{v_l+v_x}\right]F_2(x,Q^2) \nn\\
&&+\frac{x}{2}\frac{1}{v_x^2(v_l^2-v_x^2)} \left[x^2 \frac{3+9v_x+4v_x^2}{\tau\,(1+v_x)^3}-\frac{3+9v_l+4v_l^2}{\tau_l\,(1+v_l)^3}\right]\,F_1(x,Q^2)\Bigg\},\eqlab{NEWinelasticpol}\\
\Delta  E^{\prime \, \mathrm{(subtr.)} }_{nS} &=&\frac{2\al m}{\pi} \phi_n^2 \int_0^\infty \frac{\dd Q}{Q^3}\frac{2+v_l}{(1+v_l)^2}\,\ol T_1(iQ,Q^2)\eqlab{T1iQContr}.
\eea
\end{subequations}
Of course the sum of the two, i.e., the polarizability contribution, remains to be the same,
\beq
\Delta  E^\text{(pol.)}_{nS}= \Delta  E^\text{(inel.)}_{nS} + \Delta  E^\text{(subtr.)}_{nS} = \Delta  E^{\prime \, (\mathrm{inel.})}_{nS} +
\Delta  E^{\prime \, \mathrm{(subtr.)} }_{nS}.
\eeq
What changes is the relative size of the two contributions: while for
the subtraction at zero the inelastic contribution is relatively large, for the subtraction at $iQ$ it is negligible at the current level of experimental precision ($\sim 2 \,\upmu$eV). 
For example, substituting the commonly used
Bosted-Christy parametrization of the structure functions
(limited to the region $0\leq Q^2\leq 8$ GeV$^2$ and $1.1\leq W\leq 3.1$ GeV) \cite{Christy:2011}, we obtain:\footnote{The full data-driven evaluation in the dispersive approach, including also contributions from higher $Q
^2$ and larger $W$, yields \cite{Carlson:2011zd}:
$\Delta E^\text{(inel.)}_{2S}=-12.7(5)\,\upmu\text{eV}.$ Hence,  the size of high-energy contributions is of order $0.4  \, \upmu\text{eV}$, i.e.,  negligible at the current level of precision. Note that in the recent Cottingham-formula evaluation of the isospin breaking in the nucleon mass \cite{Gasser:2020mzy} these high-energy contributions are, at contrary, most important.}
\beq 
\eqlab{BCinel}
\Delta  E^{(\mathrm{inel.})}_{2S}\simeq  -12.3\,\upmu\text{eV}, \quad \mbox{versus} \quad \Delta  E^{\prime \, (\mathrm{inel.})}_{2S} \simeq 1.6\,\upmu\text{eV}.
\eeq

\begin{figure}[t]
\centering
       \includegraphics[width=9.5cm]{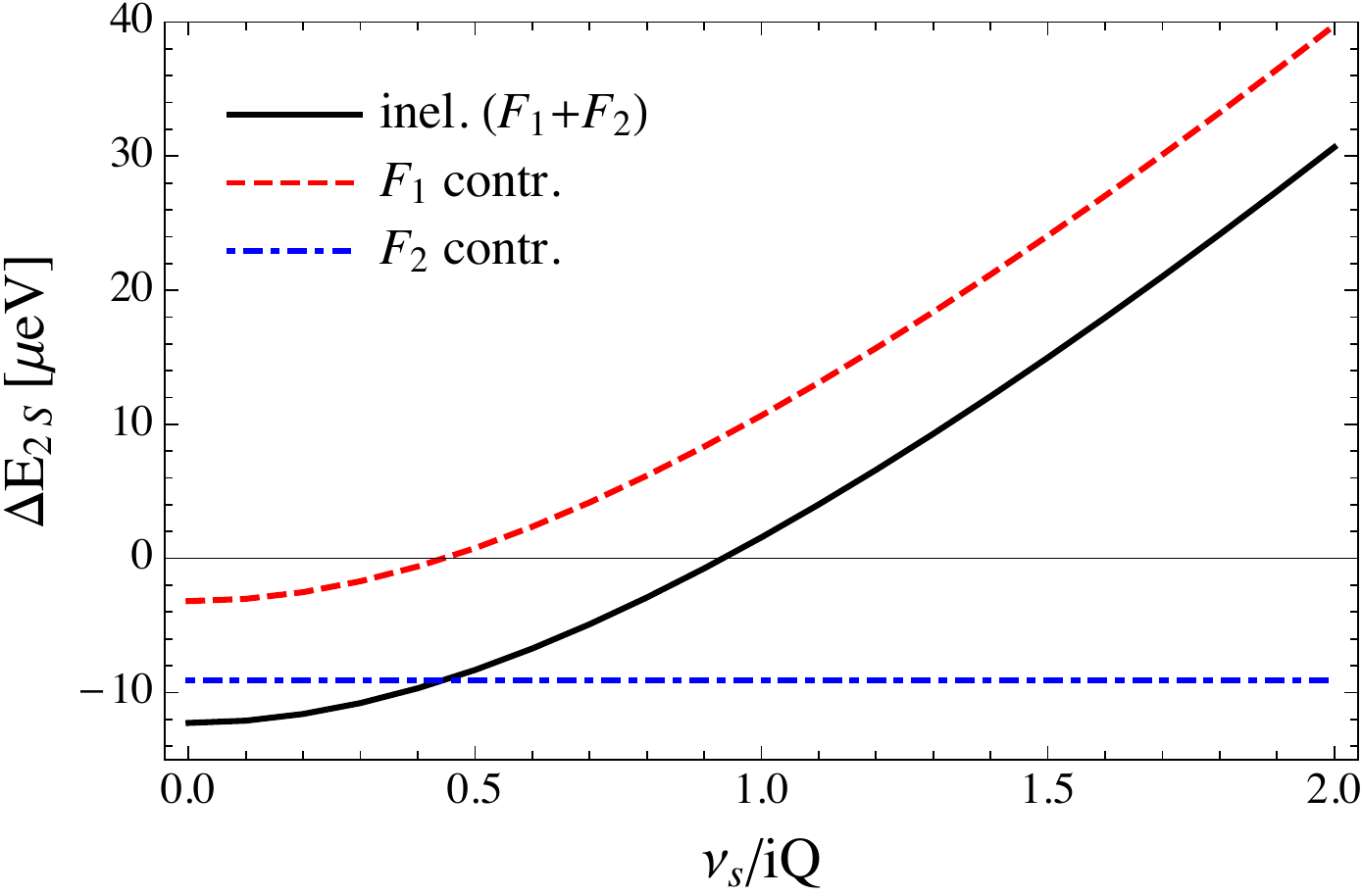}
\caption{Inelastic contribution to the $\mu$H Lamb shift as function of an arbitrary subtraction point $\nu_s$.  \label{fig:aS}}
\end{figure}

For an arbitrary subtraction point $\nu_s$, the situation is illustrated in Fig.~\ref{fig:aS}, where the inelastic contribution, along with the separate
contributions of $F_1$ and $F_2$ is considered as 
function of $\nu_s/iQ$. The values shown in \Eqref{BCinel}
correspond with the black curve at 0 and 1 respectively. One can see
that the inelastic contribution vanishes in the vicinity of $\nu_s = i Q$.
This happens for a combination of reasons, but to give a simple explanation 
consider the following.  

The TPE effect
in the Lamb shift is dominated by longitudinal photons, and as a result
by the electric form factor and the electric polarizability. The
conventional subtraction function  is, however, dominated
by transverse photons and the magnetic polarizability, 
$\ol T_1(0,Q^2) = 4\pi \beta_{M1} Q^2 + O(Q^4) $. The dominant effect is
therefore contained in the inelastic contribution. For the subtraction
at $iQ$, the subtraction function is equivalent to the purely longitudinal
amplitude, defined as $T_L = -T_1 + (1+ \nu^2/Q^2) T_2$, which at low 
$Q^2$ is given by the electric polarizability:
\beq 
\ol T_L(iQ ,Q^2)= - \ol T_1(iQ,Q^2) = 4\pi \alpha_{E1} Q^2 + O(Q^4).
\eeq 
At the same time, the integrand of the inelastic contribution is, at low $Q$, given by the longitudinal function, $F_L = - 2 x  F_1 + F_2 $.  This particular combination of structure functions is suppressed at low $Q$, due to gauge invariance. As the result, the dominant contribution is contained solely in the subtraction function. 

Concerning the transverse contributions, note that for the conventional subtraction at 0 they
cancel between the subtraction and inelastic contributions (e.g., the magnetic polarizability effect), whereas for the subtraction at $iQ$ they
are absent form the subtraction function and cancel between the structure functions within the
inelastic contribution.

\section{Prospects for lattice calculations}
The subtraction point at $iQ$ brings
a few advantages for a LQCD evaluation of the polarizability effect. 
First LQCD calculations  of the nucleon VVCS \cite{Can:2020,Hannaford-Gunn:2020pvu,Chambers:2017dov}, employing the
Feynman-Hellmann theorem \cite{Ji:2001wha}, isolate $T_1(\nu, Q^2)$ in the unphysical region. They obtain results for $Q^2\in\{ 3 \dots 7\}$ GeV$^2$ \cite{Can:2020}, with the main aim 
to compute the moments of 
$F_1$, which is done by extrapolating to $\nu=0$. The calculation of the  Lamb-shift 
contribution at the point $\nu=iQ$, proposed here, simply means setting the three-momentum of external photon to zero, $\vec q=0$.
This point can be accessed directly in lattice calculations.

The second advantage is that the evaluation at the $iQ$ point ensures that
the subtraction function
gives nearly the entire effect; the impact of the empirical structure function contribution is minimal. This said, it would be interesting
to test the lattice calculations of this effect
with structure functions. This can for instance be done by calculation the VVCS amplitude $T_1$ at
two different subtraction points. The difference
can then be expressed through an integral of the structure function $F_1$. For example, the difference between the subtraction functions 
at 0 and $iQ$ points is given by
\beq 
\ol T_1(0,Q^2)-\ol T_1(iQ,Q^2)=\frac{8\pi \al}{M}\int_0^{x_0}\dd x\, \frac{x}{\tau+x^2}F_1(x,Q^2).
\eeq

\section{Conclusions}

We have considered the possibility of an \emph{ab initio} calculation of the proton-polarizability contribution to the Lamb shift of hydrogen-like atoms. In a data-driven approach, one relies
on the dispersion relations that determine this contribution in terms of empirical structure functions of the proton, albeit not entirely. 
The subtraction function, needed for a well-defined dispersion relation involving the structure function $F_1$, is not determined experimentally and is modeled. Present state-of-art dispersive calculations model
the subtraction function, whereas the structure-function contributions are calculated by a double integration (over x and $Q^2$) of the empirical parametrizations of $F_1$ and $F_2$.
Each piece of this splitting is affected by a large Delta-resonance contribution which must cancel in the total. The empirical and modeled pieces do not have exactly the same
Delta-resonance physics, which makes the cancellation difficult to achieve in practice.

Ideally, the subtraction function should be calculated 
from LQCD. However, the problem of Delta-resonance cancellation would remain, because of the different systematics of the lattice versus empirical
evaluation. In addition, the strict zero-energy limit is 
not  directly accessible in lattice calculations. 
Here we have 
addressed both of these problems by considering a different subtraction point: $\nu=iQ$.  

Besides
the easier access of $T_1(iQ,Q^2)$  in Euclidean finite-volume calculations,
this choice has the other advantage: the polarizability contribution is  dominated by the subtraction contribution; the
structure-function contribution becomes suppressed, avoiding the large cancellations between these two contributions. Thus, while the subtraction function is directly accessed on the lattice, the structure function
contribution is dramatically reduced and hence can be calculated to better precision, relative to the full contribution.


\section*{Acknowledgements}

This work was supported by the Swiss National Science Foundation (SNSF) through the Ambizione Grant PZ00P2\_193383 and the Deutsche Forschungsgemeinschaft (DFG) through the Collaborative Research Center 1044 [The Low-Energy Frontier of the Standard Model].

  \small
\bibliography{lowQ}

\end{document}